\documentclass[aps,prapplied,twocolumn,10pt,groupedaddress,superscriptaddress,amsmath,amssymb,floatfix,showkeys]{revtex4-2}

\usepackage{graphicx}% Include figure files
\usepackage{bm,amsmath,amssymb,mathrsfs,dcolumn}
\usepackage[colorlinks=true, linkcolor=blue, citecolor=blue, urlcolor=blue, linktoc=page, bookmarks=false]{hyperref}
\usepackage{xurl} % 或者 \usepackage[hyphens]{url}
\hypersetup{breaklinks=true}
\usepackage[figure]{hypcap}
\usepackage{enumitem}
\renewcommand{\selectlanguage}[1]{}

\begin{document}
\title{Engineering-Oriented Design of Drift-Resilient MTJ Random Number Generator\\ via Hybrid Control Strategies}

\author{Ran Zhang}
    \affiliation{Beijing National Laboratory for Condensed Matter Physics, Institute of Physics,\\ University of Chinese Academy of Sciences, Chinese Academy of Sciences, Beijing 100190, China}
\author{Caihua Wan}
    \email{wancaihua@iphy.ac.cn}
    \affiliation{Beijing National Laboratory for Condensed Matter Physics, Institute of Physics,\\ University of Chinese Academy of Sciences, Chinese Academy of Sciences, Beijing 100190, China}
    \affiliation{Center of Materials Science and Optoelectronics Engineering, University of Chinese Academy of Sciences, Beijing 100049, China}
\author{Yingqian Xu}
    \affiliation{Beijing National Laboratory for Condensed Matter Physics, Institute of Physics,\\ University of Chinese Academy of Sciences, Chinese Academy of Sciences, Beijing 100190, China}
\author{Xiaohan Li}
    \affiliation{Beijing National Laboratory for Condensed Matter Physics, Institute of Physics,\\ University of Chinese Academy of Sciences, Chinese Academy of Sciences, Beijing 100190, China}
\author{Raik Hoffmann}
    \affiliation{Fraunhofer IPMS, Center Nanoelectronic Technologies, 01109 Dresden, Germany}
\author{Meike Hindenberg}
    \affiliation{Fraunhofer IPMS, Center Nanoelectronic Technologies, 01109 Dresden, Germany}
\author{Shiqiang Liu}
    \affiliation{Beijing National Laboratory for Condensed Matter Physics, Institute of Physics,\\ University of Chinese Academy of Sciences, Chinese Academy of Sciences, Beijing 100190, China}
\author{Dehao Kong}
    \affiliation{Beijing National Laboratory for Condensed Matter Physics, Institute of Physics,\\ University of Chinese Academy of Sciences, Chinese Academy of Sciences, Beijing 100190, China}
\author{Shilong Xiong}
    \affiliation{Beijing National Laboratory for Condensed Matter Physics, Institute of Physics,\\ University of Chinese Academy of Sciences, Chinese Academy of Sciences, Beijing 100190, China}
\author{Shikun He}
    \affiliation{Zhejiang Hikstor Technology Co. Ltd, Hangzhou 311305, China}
\author{Alptekin Vardar}
    \affiliation{Fraunhofer IPMS, Center Nanoelectronic Technologies, 01109 Dresden, Germany}
\author{Qiang Dai}
    \affiliation{Zhejiang Hikstor Technology Co. Ltd, Hangzhou 311305, China}
\author{Junlu Gong}
    \affiliation{Zhejiang Hikstor Technology Co. Ltd, Hangzhou 311305, China}
\author{Yihui Sun}
    \affiliation{Zhejiang Hikstor Technology Co. Ltd, Hangzhou 311305, China}
\author{Zejie Zheng}
    \affiliation{Zhejiang Hikstor Technology Co. Ltd, Hangzhou 311305, China}
\author{Thomas Kämpfe}
    \email{thomas.kaempfe@ipms.fraunhofer.de}
    \affiliation{Fraunhofer IPMS, Center Nanoelectronic Technologies, 01109 Dresden, Germany}
    \affiliation{TU Braunschweig, Institute for CMOS Design, 38106 Braunschweig, Germany}
\author{Guoqiang Yu}
    \affiliation{Beijing National Laboratory for Condensed Matter Physics, Institute of Physics,\\ University of Chinese Academy of Sciences, Chinese Academy of Sciences, Beijing 100190, China}
    \affiliation{Center of Materials Science and Optoelectronics Engineering, University of Chinese Academy of Sciences, Beijing 100049, China}
    \affiliation{Songshan Lake Materials Laboratory, Dongguan, Guangdong 523808, China}
\author{Xiufeng Han}
    \email{xfhan@iphy.ac.cn}
    \affiliation{Beijing National Laboratory for Condensed Matter Physics, Institute of Physics,\\ University of Chinese Academy of Sciences, Chinese Academy of Sciences, Beijing 100190, China}
    \affiliation{Center of Materials Science and Optoelectronics Engineering, University of Chinese Academy of Sciences, Beijing 100049, China}
    \affiliation{Songshan Lake Materials Laboratory, Dongguan, Guangdong 523808, China}

\date{\today}% It is always \today, today,

\begin{abstract}
  Magnetic Tunnel Junctions (MTJs) have shown great promise as hardware sources for true random number generation (TRNG) due to their intrinsic stochastic switching behavior. However, practical deployment remains challenged by drift in switching probability caused by thermal fluctuations, device aging, and environmental instability. This work presents an engineering-oriented, drift-resilient MTJ-based TRNG architecture, enabled by a hybrid control strategy that combines self-stabilizing feedback with pulse width modulation. A key component is the Downcalibration-2 scheme, which updates the control parameter every two steps using only integer-resolution timing, ensuring excellent statistical quality without requiring bit discarding, pre-characterization, or external calibration. Extensive experimental measurements and numerical simulations demonstrate that this approach maintains stable randomness under dynamic temperature drift, using only simple digital logic. The proposed architecture offers high throughput, robustness, and scalability, making it well-suited for secure hardware applications, embedded systems, and edge computing environments.
\end{abstract}
\maketitle

\begingroup
\renewcommand\thefootnote{}
\footnotetext{* Corresponding authors: wancaihua@iphy.ac.cn, thomas.kaempfe@ipms.fraunhofer.de, xfhan@iphy.ac.cn}
\endgroup

\section{Introduction}

In the digital age, the quest for secure and reliable random number generation has become a cornerstone of modern technology. As our world becomes increasingly interconnected, the demand for robust cryptographic systems \cite{choi_volatile_2024}, secure communications \cite{lee_trng_2018}, and sophisticated simulations \cite{ferrenberg_monte_1992} has surged. At the heart of these systems lies the need for true randomness—a quality that ensures unpredictability and security in digital processes. Random numbers are the bedrock of encryption algorithms, safeguarding sensitive information from unauthorized access and ensuring the integrity of data transmission. They are also crucial in simulations, where they enable the modeling of complex systems and phenomena with high fidelity. The importance of random number generation extends beyond traditional computing applications. In fields such as finance \cite{joy_quasi-monte_1996}, healthcare \cite{stout_keeping_2008}, and artificial intelligence \cite{li_memory_2022, li_restricted_2024}, the ability to generate truly random numbers is essential for risk assessment \cite{barry_recommendations_1996}, decision-making \cite{si_energy-efficient_2024, zhang_probabilistic_2025}, and the development of intelligent systems \cite{grollier_neuromorphic_2020, markovic_physics_2020, mohseni_ising_2022, finocchio_roadmap_2024}. As these domains continue to evolve, the demand for high-quality random numbers grows, driving the need for innovative solutions that can meet the stringent requirements of modern applications \cite{camsari_stochastic_2017, camsari_p-bits_2019}.

Historically, random number generation has relied on deterministic algorithms, known as pseudo-random number generators (PRNGs) \cite{wichmann_algorithm_1982, blakley_efficient_2000}, which, while efficient, lack the true randomness required for high-security applications. The limitations of PRNGs have spurred the exploration of alternative technologies capable of producing genuine randomness. Among these, True Random Number Generators (TRNGs) have emerged as a promising solution, leveraging physical processes to generate random numbers that are inherently unpredictable.

In the pursuit of effective TRNGs, researchers have explored a variety of physical phenomena, from thermal noise and quantum effects to chaotic systems. Each approach offers unique advantages and challenges, with the ultimate goal of achieving a balance between randomness quality, speed, and integration capability. It is within this context of innovation and exploration that Magnetic Tunnel Junctions (MTJs) have gained attention as a viable technology for TRNGs \cite{lee_design_2017, vodenicarevic_low-energy_2017, liu_spin_2018, chen_binary_2018, li_spin_2018, qin_thermally_2019, fukushima_recent_2021, song_power_2021, zhang_time_2021, debashis_gaussian_2022, li_stochastic_2023, li_true_2023, singh_cmos_2024, gibeault_programmable_2024,  xu_self-stabilized_2024, luo_magnetic_2024, zhang_probabilitydistributionconfigurable_2024, bao_computing_2024}. Their potential to harness the stochastic nature of magnetization switching presents a novel avenue for generating true random numbers, offering a promising alternative to traditional methods \cite{borders_integer_2019, misra_probabilistic_2022}.

MTJs, particularly those utilizing spin-transfer torque (STT) mechanisms \cite{myers_current-induced_1999, krivorotov_time-domain_2005, devolder_single-shot_2008, ralph_spin_2007, fukushima_spin_2014,hong_demonstration_2019, rehm_stochastic_2023, dubovski_one_2024}, offer a novel approach to random number generation by exploiting the stochastic nature of magnetization switching. The fundamental operation of an MTJ-based TRNG relies on the ability to toggle between distinct resistance states—parallel (P) and antiparallel (AP)—which correspond to binary values \cite{zhao_type-y_2022,liu_magnetization_2024}. This switching is driven by voltage pulses, and the inherent randomness in the switching process is harnessed to produce random bits \cite{camsari_implementing_2017, zhang_tunable_2021, camsari_double-free-layer_2021, chen_magnetic-tunnel-junction-based_2022}. The high tunneling magnetoresistance (TMR) ratio of MTJs, often exceeding 200\%, provides a clear distinction between the resistance states, facilitating reliable detection and interpretation of the generated bits \cite{camsari_charge_2020, rehm_stochastic_2023, valli_high-speed_2025}. The TMR ratio is defined as \( \text{TMR} = \frac{R_{\text{AP}} - R_{\text{P}}}{R_{\text{P}}} \times 100\% \), where \( R_{\text{AP}} \) and \( R_{\text{P}} \) represent the resistance in the AP and P states, respectively.

However, the practical implementation of MTJ-based TRNGs is not without challenges. A significant issue is the drift in switching probability over time, which can compromise the randomness and bias the output. This drift is influenced by several factors, including the training effect \cite{zhao_failure_2016} and the heating effect \cite{lv_seebeck_2022} of magnetic materials induced by current flow during switching events. These phenomena can alter the magnetic properties of the MTJ, leading to a gradual shift in the switching probability and, consequently, the randomness of the generated numbers \cite{xu_self-stabilized_2024}. Moreover, the temperature dependence of the switching probability adds another layer of complexity, as fluctuations in ambient temperature can significantly impact the voltage required to achieve a specific switching probability \cite{rehm_temperature-resilient_2024}.

Addressing these challenges requires innovative control strategies that can stabilize the switching probability and mitigate the effects of drift. In this context, we propose a hybrid control strategy that combines self-stabilization techniques with pulse width modulation to enhance the performance of MTJ-based TRNGs. The self-stabilization strategy dynamically adjusts the driving voltage based on real-time feedback from the MTJ's resistance state, ensuring that the switching probability remains close to the target value. This approach leverages the continuous and monotonic relationship between voltage and switching probability, allowing for precise control without the need for resource-intensive probability measurements. Furthermore, we explore the potential of pulse width modulation as an alternative method for tuning the switching probability. By analyzing the dependence of switching probability on both voltage amplitude and pulse width, we demonstrate that pulse width offers a practical and efficient means of control, particularly in systems where fine voltage adjustments are challenging. The insights gained from this analysis inform the design of a pulse-width-based tunable TRNG, which utilizes an 8-bit control word to achieve precise pulse width adjustments and produce a well-defined probability distribution.

This paper presents a comprehensive study of the engineering-oriented design of drift-resilient MTJ-based TRNGs, highlighting the effectiveness of hybrid control strategies in overcoming the limitations of traditional approaches. Through detailed experimental and simulation results, we demonstrate the improved stability, uniformity, and quality of the random numbers generated by our proposed system. The findings underscore the potential of MTJ-based TRNGs as a reliable solution for applications requiring high-quality randomness, offering flexibility and adaptability for diverse use cases in cryptography, secure communications, and beyond.

\section{Methods}

\begin{figure*}
\includegraphics[width=\textwidth]{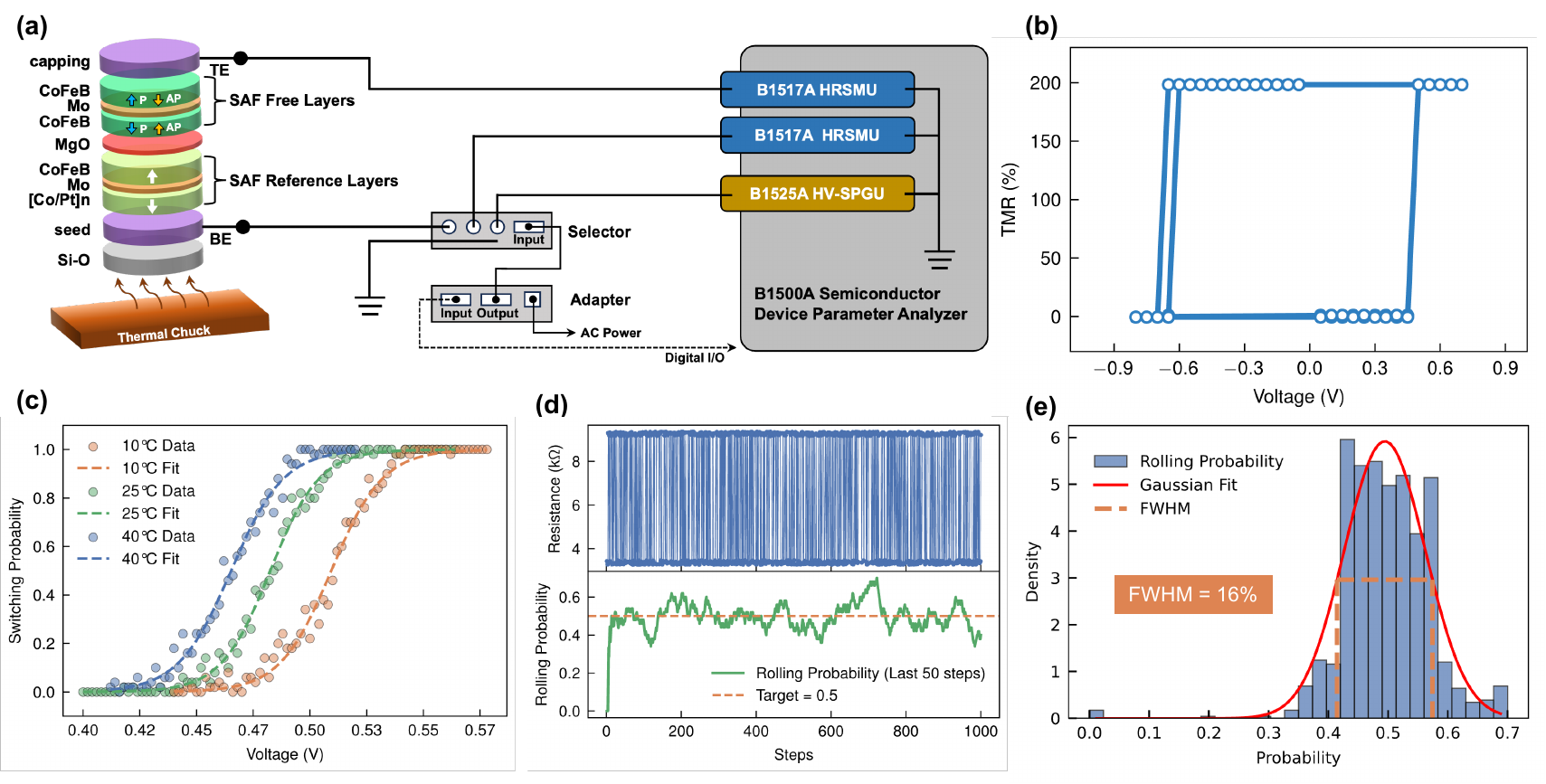}
\caption{\label{fig:device}Drift issue in MTJ-based true random number generators. (a) Schematic representation of the device structure and measurement setup. The MTJ stack consists of a thermal chuck for precise temperature control and a layered structure, including a reference layer, a free layer, and an antiferromagnetic layer. Electrical connections facilitate input and output signals for resistance measurement. The testing configuration utilizes the B1500A Semiconductor Device Parameter Analyzer with two high-resolution source measurement units (B1517A HRSMU) and one high-voltage source/pulse generator unit (B1525A HV-SPGU). (b) Voltage-driven resistance switching behavior of the MTJ under a typical 50 ns pulse duration, corresponding to magnetization switching of the free layer. The resistance states reflect the parallel (P) and antiparallel (AP) magnetic configurations, characterized by a tunneling magnetoresistance (TMR) ratio of approximately 200\%. (c) Switching probability of the MTJ as a function of voltage for different operating temperatures (10°C, 25°C, and 40°C). The data points and corresponding fits demonstrate a temperature-dependent shift in the voltage required to achieve a specific switching probability. The three curves from left to right correspond to measurements taken at 10°C, 25°C, and 40°C, respectively. (d) The output of the random bitstream generated by the MTJ (targeting a 50\% switching probability). The top panel shows the resistance states as a function of time, and the lower panel presents the rolling probability (calculated over the last 50 steps). (e) Probability distribution of the rolling probability. The histogram exhibits a crude Gaussian distribution, with the solid red line representing the Gaussian fit and the dashed orange line indicating the full width at half maximum (FWHM) of the fit, approximately 16\%.}
\end{figure*}

The STT-MTJs were fabricated through magnetron sputtering deposition under high vacuum conditions (\(10^{-6}\) Pa). The multilayer stack structure consists of capping/CoFeB/Mo/CoFeB/MgO/CoFeB/Mo/[Co/Pt]\(_n\)-based synthetic antiferromagnetic layers deposited (top to bottom) on thermally oxidized silicon substrates (FIG.~\hyperref[fig:device]{\ref{fig:device}(a)}). Post-deposition thermal annealing was conducted at elevated temperature under an out-of-plane magnetic field. Standard photolithography and etching processes were employed to pattern the films into cylindrical STT-MTJ devices with an approximate diameter of 40 nm. The magneto-transport characterization was performed using an Hprobe H3DM tester. Electrical measurements were carried out by connecting the devices to a Keysight B1500A semiconductor analyzer through a probe card setup as shown in FIG.~\hyperref[fig:device]{\ref{fig:device}(a)}. The write pulses applied during switching measurements had a typical duration of approximately 50 ns, unless otherwise specified. A Python-based interface enabled automated control and data acquisition for systematic evaluation of switching probabilities, establishing a reliable platform for assessing the TRNG performance of the STT-MTJs.

\section{Results and Discussion}

\subsection{Self-stablization Strategies}

\subsubsection{Drift Issue in MTJ-based TRNG}

The MTJ's switching behavior is governed by the voltage applied to the device. As shown in FIG.~\hyperref[fig:device]{\ref{fig:device}(b)}, the MTJ can be switched between two distinct resistance states: P and AP. These states correspond to different orientations of the magnetization in the free and reference layers, characterized by a high TMR ratio of approximately 200\%. The low resistance state (P) represents a "1" bit, while the high resistance state (AP) represents a "0" bit. The resistance state of the MTJ is determined by the voltage pulse applied to the device, driving the magnetization of the free layer to switch between these two states. The MTJ's switching behavior is highly sensitive to the applied voltage, and this voltage-driven switching is essential for generating random numbers. The ability to reliably induce and detect these switching events is critical for achieving unbiased random number generation.

However, the switching characteristics of MTJs are highly sensitive to external and internal factors, particularly temperature variations, as shown in FIG.~\hyperref[fig:device]{\ref{fig:device}(c)}. This figure illustrates the switching probability of the MTJ as a function of voltage at different operating temperatures (10°C, 25°C, and 40°C). A clear shift in the required voltage for achieving a specific switching probability is evident as the temperature increases. For example, at elevated temperatures such as 40°C, the thermal energy reduces the effective energy barrier, thus lowering the voltage needed for switching. This temperature dependence demonstrates the sensitivity of the MTJ's critical switching conditions to external conditions, posing challenges for the stability and reliability of random number generation in practical environments.

FIG.~\hyperref[fig:device]{\ref{fig:device}(d)} further evaluates the performance of the MTJ-based TRNG under continuous sampling conditions, targeting a switching probability of 50\%. The top panel shows the switching between the parallel and antiparallel resistance states over time, and the bottom panel displays the rolling switching probability calculated over the last 50 steps. While the rolling probability exhibits noticeable fluctuations, it should be noted that this fluctuation alone does not directly imply long-term drift of the probability. Instead, these short-term fluctuations are inherent statistical variations due to the limited sampling window size.

To quantitatively assess the nature of these fluctuations, the probability distribution of the rolling probability is analyzed, as shown in FIG.~\hyperref[fig:device]{\ref{fig:device}(e)}. The histogram presents the distribution of the rolling probability, exhibiting a crude Gaussian distribution. The solid red line represents the Gaussian fit to the data, while the dashed orange line indicates the full width at half maximum (FWHM) of approximately 16\%. Notably, this FWHM value aligns with the theoretical expectation for a 50-step rolling window,
\begin{equation}
\mathrm{FWHM} \approx 2.355 \times \sqrt{\frac{0.5 \times 0.5}{50}} = 16.5\%
\end{equation}
indicating that the width of the observed distribution primarily results from statistical fluctuations inherent to the binomial process itself, rather than from any approximation. Notably, the standard deviation formula used here, $\sigma = \sqrt{Np(1-p)}$, is the exact result for the binomial distribution, not a Gaussian approximation. However, the coarse quality of the Gaussian fit reveals underlying imperfections in randomness, suggesting the existence of subtle non-random features, such as correlations or systematic biases within the bitstream.

\subsubsection{Step-by-Step Calibration}

\begin{figure*}
\includegraphics[width=\textwidth]{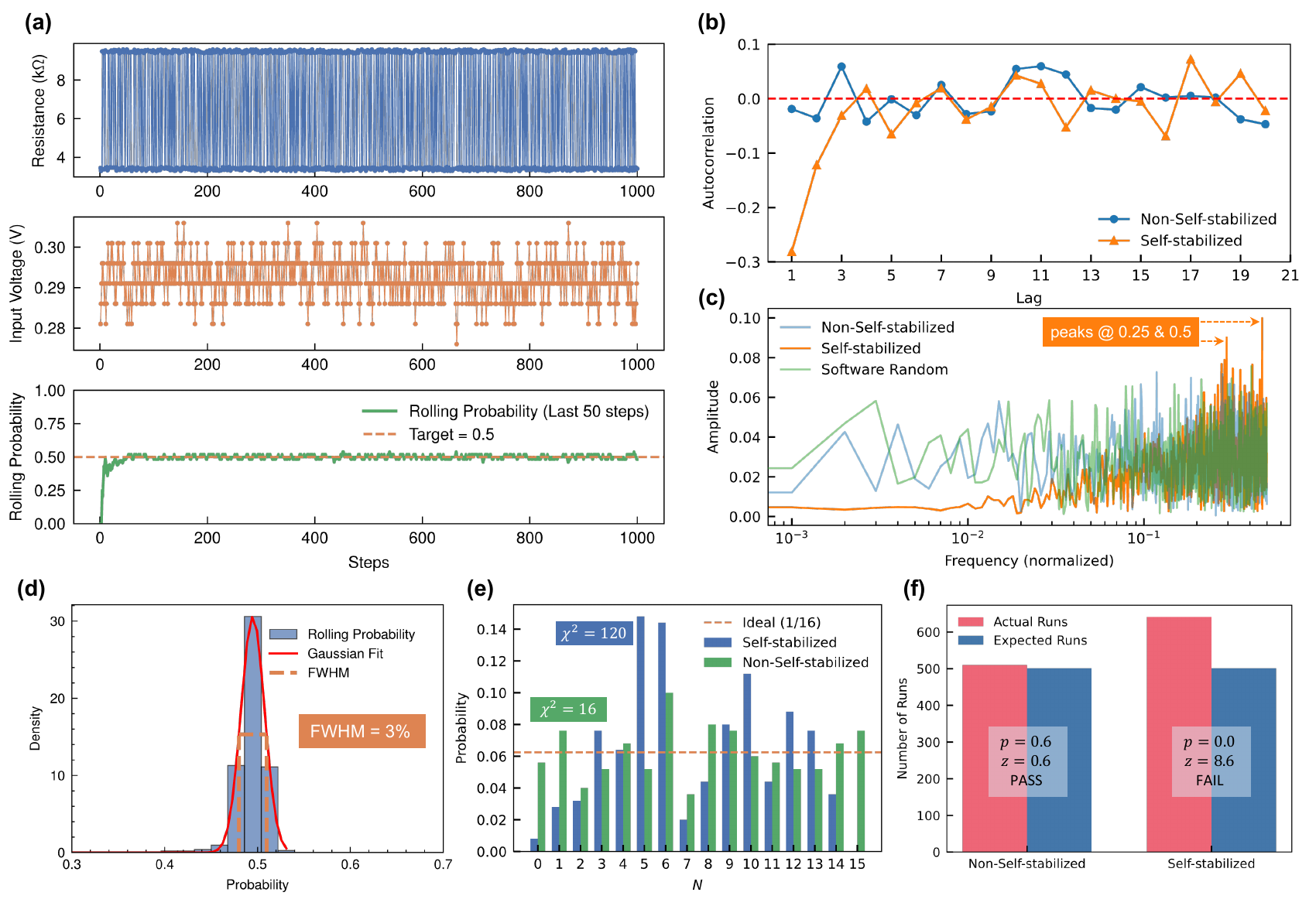}
\caption{\label{fig:self}Results of the step-by-step self-stabilizing strategy for MTJ-based true random number generators. (a) Output targeting a switching probability of 50\%. The top panel shows the resistance switching behavior over time, the middle panel presents the dynamically adjusted driving voltage of the MTJ during each calibration step, and the bottom panel illustrates the rolling probability (calculated over the last 50 steps), converging to the target value of 0.5. (b) Comparison of autocorrelation coefficients, showing that the self-stabilized strategy exhibits significantly higher autocorrelation at lag 1 and 2. (c) Comparison of FFT spectra, where the self-stabilized strategy clearly shows distinct spectral features compared to the non-stabilized case. (d) Gaussian distribution fit of the rolling probability under the self-stabilized strategy. The FWHM of the distribution is approximately 3\%, indicating compressed randomness and potential bit-to-bit correlation. (e) 4-bit chi-square uniformity test under the self-stabilized strategy. The distribution is analyzed by combining four adjacent Bernoulli random variables $A$, $B$, $C$, and $D$ into a decimal number $N$ ranging from 0 to 15, calculated as \(N=8A+4B+2C+D\). The self-stabilized TRNG exhibits worse uniformity compared to the non-calibrated system, as evidenced by a significantly larger chi-square value. (f) Comparison of the run test results, showing that the self-stabilized strategy fails the test with a notably higher number of runs, suggesting increased 01–10 correlation.}
\end{figure*}

To address the drift problem commonly observed in MTJ-based TRNGs, we proposed a step-by-step self-stabilization strategy \cite{xu_self-stabilized_2024}. The main idea is to circumvent the need for direct probability measurements by leveraging the monotonic relationship between the write voltage and the switching probability. This relationship allows us to treat voltage as a controllable proxy for probability, enabling real-time correction of switching behavior without the need for statistical post-processing.

The step-by-step calibration strategy operates iteratively: after each switching event, the resistance state of the MTJ is read and interpreted as a binary outcome ("1" or "0"). Based on this result and the desired target switching probability \(P_0\), the write voltage is incrementally adjusted—reduced if a "1" is observed, or increased if a "0" is observed. The voltage is updated according to a simple rule:

\setlist[itemize]{left=0pt}
\begin{itemize}
    \item If the output bit is \(1\), set the new voltage to \(V - (1 - P_0)\Delta V\);
    \item If the output bit is \(0\), set the new voltage to \(V + P_0\Delta V\),
\end{itemize}where \(\Delta V\) is a small, fixed voltage step size. This feedback loop allows the switching probability to rapidly converge to the target value, as shown in FIG.~\hyperref[fig:self]{\ref{fig:self}(a)}. In the bottom panel of the figure, the rolling probability over the last 50 steps clearly stabilizes at 0.5, indicating the efficacy of the strategy in suppressing drift. However, despite this apparent success in achieving statistical stability over short windows, a closer inspection reveals a critical limitation of the step-by-step self-stabilization method: it introduces strong bit-to-bit correlations, which degrade the overall randomness quality of the generated bitstream. This limitation becomes evident when analyzing deeper statistical properties of the output.

As shown in FIG.~\hyperref[fig:self]{\ref{fig:self}(b)}, the autocorrelation coefficients at lag 1 and lag 2 are significantly elevated in the self-stabilized case, in stark contrast to the nearly uncorrelated behavior of the non-stabilized reference. These elevated autocorrelation values imply that consecutive bits are no longer independent, and that the outcome of a given bit is influenced by the outcome of the preceding bits. This violates a fundamental requirement for TRNGs: the statistical independence of each output bit. Further frequency-domain analysis, illustrated in FIG.~\hyperref[fig:self]{\ref{fig:self}(c)}, supports this conclusion. The fast Fourier transform (FFT) spectrum of the self-stabilized output displays distinct periodic components that are absent in the non-stabilized output. Such features are indicative of quasi-deterministic structures embedded within the output bitstream, and they compromise its suitability for applications requiring high entropy. The Gaussian distribution of the rolling probability in FIG.~\hyperref[fig:self]{\ref{fig:self}(d)} at first appears to confirm high stability, with a FWHM of approximately 3\%, significantly narrower than the 16\% FWHM observed in the unregulated system (cf. FIG.~\hyperref[fig:device]{\ref{fig:device}(e)}). However, this narrow distribution actually reflects a different issue: over-confinement of the switching behavior, which suppresses natural randomness. The distribution is not only narrow but also coarsely shaped, deviating from the smooth, statistically expected Gaussian form. This suggests that the random process is being overly constrained by the control loop, reducing entropy and introducing predictability.

This issue is further quantified by a 4-bit chi-square uniformity test, shown in FIG.~\hyperref[fig:self]{\ref{fig:self}(e)}. By grouping four consecutive bits into a 4-bit integer \(N = 8A + 4B + 2C + D\), we construct a histogram over all 16 possible values. Ideally, the histogram should approach a uniform distribution if the output bits are statistically independent and identically distributed. However, the self-stabilized output yields a highly uneven distribution with a large chi-square value, far exceeding that of the non-stabilized system. This is strong evidence that the step-by-step strategy compromises uniformity. Finally, in FIG.~\hyperref[fig:self]{\ref{fig:self}(f)}, the run test results further underscore the problem. The self-stabilized output fails the run test, with a significantly elevated number of alternating "01" and "10" patterns. This again points to a pronounced alternation tendency—likely a direct consequence of the feedback mechanism—which biases the generator toward switching states in a pseudo-regular fashion, rather than randomly.

In summary, while the step-by-step self-stabilization strategy effectively suppresses long-term drift and ensures rapid convergence to a target switching probability, it does so at the cost of short-range statistical independence. The method imposes strong temporal correlations that undermine the entropy and uniformity of the output bitstream. Although suitable for stabilizing low-rate randomness under tightly controlled conditions, this strategy, in its basic form, fails to meet the rigorous statistical requirements of high-quality TRNGs. These limitations motivate the development of refined strategies—such as periodic calibration or selective discarding—that preserve the benefits of voltage regulation while restoring statistical independence.

\subsubsection{Downsampling and Downcalibration}

\begin{figure*}
\includegraphics[width=\textwidth]{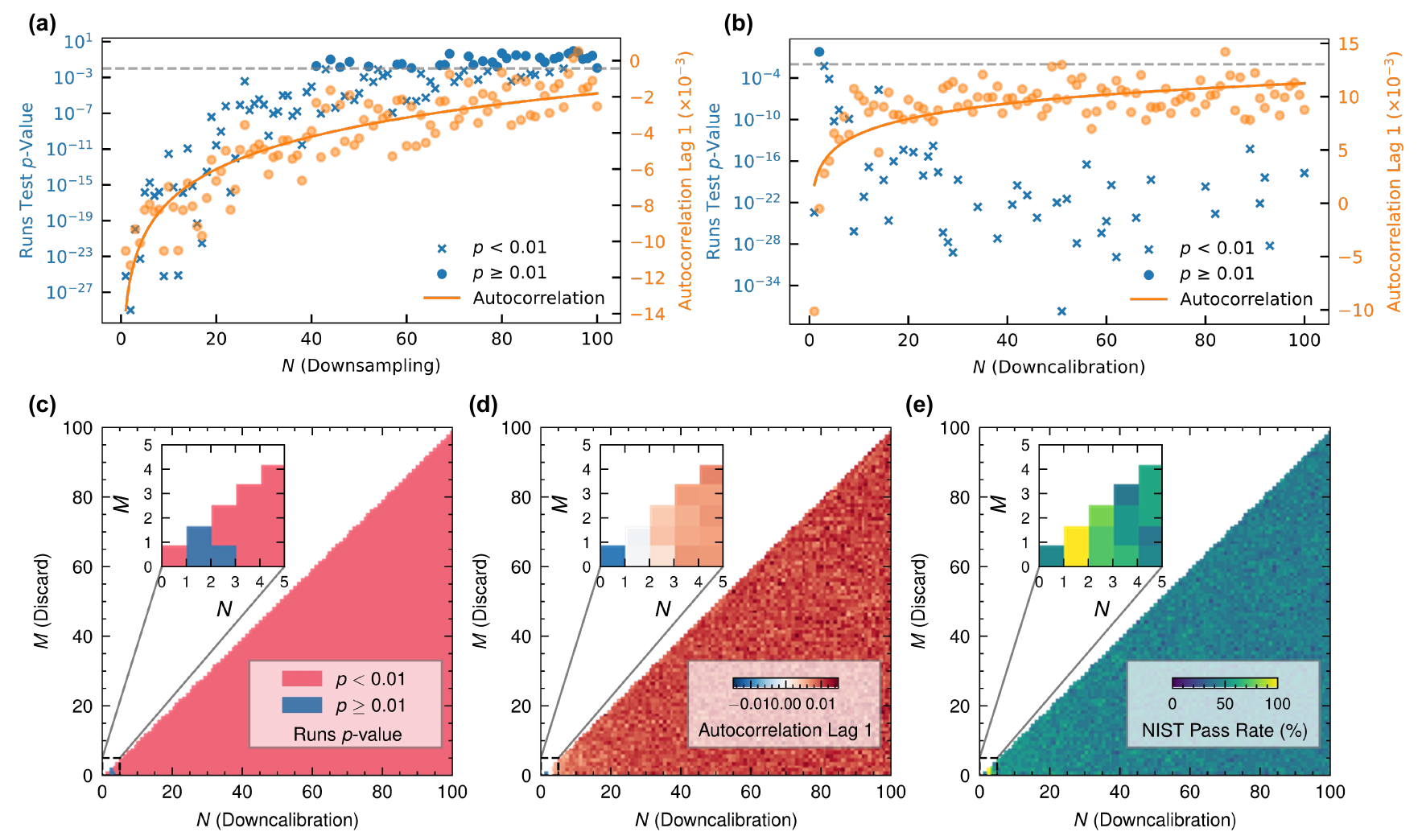}
\caption{\label{fig:nm}Evaluation of modified sampling and calibration methods for the self-stabilized strategy. (a) Results of downsampling every $N$ steps (Downsampling-$N$): the runs test $p$-values (left axis, blue) and the lag-1 autocorrelation coefficients (right axis, orange) versus $N$. The runs test occasionally passes for $N > 40$ (gray line denotes the threshold $p = 0.01$, circles indicate passed data points, crosses indicate failed ones). The absolute value of autocorrelation decreases gradually with increasing $N$. (b) Results of calibrating every $N$ steps (Downcalibration-$N$): runs test $p$-values and lag-1 autocorrelation coefficients versus $N$. The runs test only passes when $N = 2$, where the autocorrelation reaches its minimum absolute value. (c–e) Results for calibrating every $N$ steps while discarding $M$ data points after each calibration (Downcalibration-$N$ with Discard-$M$). Shown are the runs test results (c), lag-1 autocorrelation coefficients (d), and the pass rate of the 15 NIST SP 800-22 tests (e). The runs test only passes for $(N, M) = (2, 0)$, $(2, 1)$, and $(3, 0)$. Only $(N, M) = (2, 0)$ and $(2, 1)$ pass the lag-1 autocorrelation test and all 15 NIST randomness tests.}
\end{figure*}

\begin{figure}
\includegraphics[width=0.48\textwidth]{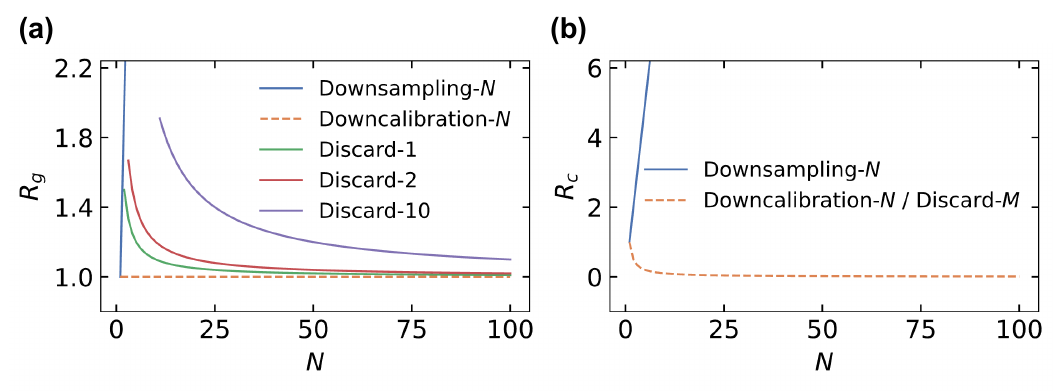}
\caption{\label{fig:rate}Theoretical analysis of effective bit rates under different downsampling and downcalibration strategies. (a) Ratio between the number of generated raw bits and the number of effective bits $R_g$. The Downcalibration-$N$ method achieves the lowest ratio, meaning no raw bits are wasted. The Downsampling-$N$ method results in the highest ratio, requiring $N$ raw bits to produce one effective bit. The Downcalibration-$N$ with Discard-$M$ strategy lies in between, with the ratio given by $1 + M/N$, increasing with $M$. (b) Ratio between the number of calibration operations and the number of effective bits $R_c$. For the Downcalibration-$N$ and Downcalibration-$N$ with Discard-$M$ methods, this ratio is $1/N$. In contrast, the Downsampling-$N$ method requires one calibration per raw bit, resulting in $N$ calibration operations per effective bit.}
\end{figure}

To overcome the limitations of the step-by-step self-stabilization strategy—particularly its tendency to introduce strong bit-to-bit correlations—we explored three modified methods aimed at improving randomness while retaining the benefits of voltage-based drift suppression. These methods are based on adjusting the timing and frequency of calibration and sampling, and are named as follows: Downsampling-$N$, Downcalibration-$N$, and Downcalibration-$N$ with Discard-$M$.

It is important to note that all results presented in this section are obtained from numerical simulations. Specifically, we implemented the MTJ-based switching model and calibration logic in Python, and used Python’s built-in software random number generator to simulate the stochastic switching behavior. Although this approach does not fully replicate the physics of real MTJs, it provides a reliable and controllable environment to evaluate algorithmic strategies and their statistical consequences under idealized conditions. The software random number generator serves as a stand-in for probabilistic switching, allowing us to focus on the effects of the control logic rather than hardware imperfections.

The first strategy, Downsampling-$N$, maintains the original feedback mechanism that adjusts the write voltage after every bit, but only keeps one out of every $N$ output bits. The idea is that by skipping intermediate bits, the influence of temporal correlations between consecutive outputs can be reduced. As shown in FIG.~\ref{fig:nm}(a), this method does show some improvement in the runs test $p$-values and a gradual reduction in autocorrelation as $N$ increases. The runs test occasionally passes for $N > 40$, though the results are inconsistent. However, this comes at a steep cost in bit efficiency. Since only one bit is retained every $N$ cycles, the number of raw bits required to generate a usable output bit is large, as confirmed by the theoretical analysis in FIG.~\ref{fig:rate}(a). Additionally, the calibration overhead remains high, as shown in FIG.~\ref{fig:rate}(b), because a voltage update is still performed for every bit, whether or not it is used. This makes Downsampling-$N$ unsuitable for high-throughput or energy-constrained applications.

The second strategy, Downcalibration-$N$, takes the opposite approach. Instead of adjusting the voltage after every bit, the system fixes the voltage for $N$ consecutive bits and only recalibrates the voltage every $N$ steps. All generated bits are retained. This method is more efficient in terms of both bit output and calibration operations, as seen in FIG.~\ref{fig:rate}(a) and \ref{fig:rate}(b), where both bit efficiency and calibration efficiency are optimal. FIG.~\ref{fig:nm}(b) shows that the runs test passes only for $N = 2$, which also corresponds to the minimum in autocorrelation. Interestingly, this configuration provides not only minimal correlation but also avoids the performance cost of discarding bits. This finding was unexpected, as one might assume that discarding is necessary to decorrelate the output. The effectiveness of Downcalibration-2 shows that it is possible to achieve high-quality randomness without any bit loss, simply by calibrating the system once every two steps.

The third method, Downcalibration-$N$ with Discard-$M$, combines the timing of the second strategy with selective bit removal. After each calibration, $M$ bits are discarded, and only the remaining $N - M$ bits are used as output. This method introduces a tunable parameter to balance randomness and efficiency. FIG.~\ref{fig:nm}(c)–(e) show the runs test results, autocorrelation coefficients, and NIST test pass rates for various combinations of $(N, M)$. Only a few parameter pairs, including $(2, 0)$, $(2, 1)$, and $(3, 0)$, pass the runs test. Among these, only $(2, 0)$ and $(2,1)$ also pass the autocorrelation test and achieve a full 15/15 pass rate in the NIST SP 800-22 test suite, as indicated by a 100\% pass rate.

The most remarkable result from this exploration is the discovery that Downcalibration-2—in which the control parameter is updated once every two bits without discarding any output—achieves near-optimal statistical performance. This strategy successfully passes the full NIST SP 800-22 test suite, exhibits minimal autocorrelation, and avoids the throughput loss associated with data discarding. Such performance was initially counterintuitive, as immediate feedback is often assumed to introduce temporal correlations that are only mitigated by discarding or heavy downsampling. To validate that this result is not an artifact of the simulation environment, we repeated all experiments using a hardware-based TRNG embedded in Apple’s M2 Pro processor. The statistical behavior remained consistent, confirming that the performance of Downcalibration-2 is robust and not dependent on the properties of the software pseudo-random number generator.

However, a deeper analysis revealed an important caveat: although Downcalibration-2 passes all NIST tests and maintains excellent short-range independence, it fails the chi-square test when analyzing the uniformity of grouped 4-bit symbols. This test is highly sensitive to subtle biases in symbol frequency and is known to fail even for widely used pseudo-random number generators such as Python’s random module or early Java Random implementations, particularly on short streams or small symbol spaces. The deviation we observe likely originates from residual structure due to periodic feedback and insufficient entropy mixing at the multi-bit level. Although this does not affect applications relying on single-bit entropy—such as Monte Carlo methods, p-computing, or cryptographic masking—it limits use in cases requiring strong uniformity in grouped outputs. We view this as an important direction for future work, including integrating entropy extractors, decorrelating encoders, or adaptive calibration schedules.

Interestingly, we propose a physical explanation for the remarkable effectiveness of Downcalibration-2, grounded in the finite precision of real-world control hardware. In practice, voltage or pulse-width adjustments are not continuous, but quantized—e.g., limited to 1 mV (or 1 ns as will be discussed below) steps. Suppose a desired switching probability of 50\% corresponds to an ideal voltage of 0.2815 V; due to resolution limits, the actual output can only be 0.281 V or 0.282 V. Either choice yields a slight offset from the true midpoint, resulting in a persistent positive correlation. The Downcalibration strategy introduces a negative correlation by design—feedback adjusts the control parameter in the opposite direction after each observed output. When applied every two steps, this negative correlation can precisely counterbalance the residual positive bias introduced by finite control resolution. If calibration occurs every three or more steps, the accumulated error cannot be fully compensated, and correlations re-emerge. In this sense, the observed optimality of Downcalibration-2 may reflect a subtle but fundamental quantization balancing mechanism, akin to digital compensation methods used in quantized neural networks or low-precision optimization. Our simulations explicitly include such quantization effects, reinforcing the practical relevance of this result: in systems where analog control is inherently limited, lightweight, discrete-time digital feedback offers an effective and scalable path toward robust and high-quality randomness.

\subsubsection{Automatic Convergence}

\begin{figure*}
\includegraphics[width=\textwidth]{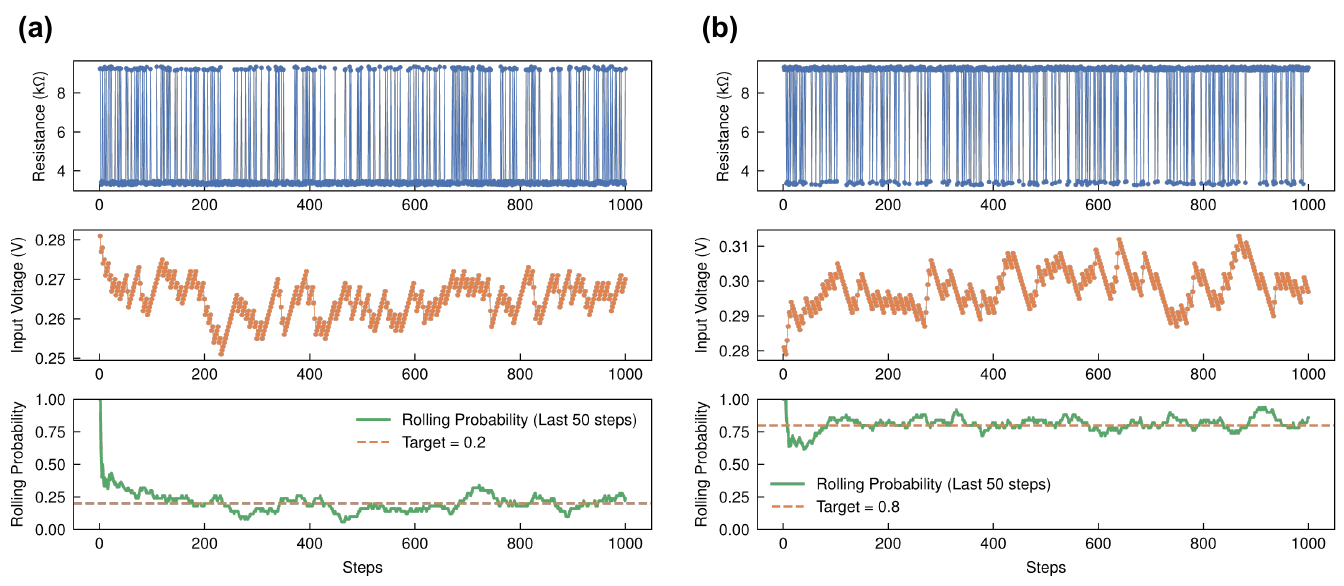}
\caption{\label{fig:twoeight}The Downcalibration-2 strategy outputs target switching probabilities of 20\% (a) and 80\% (b). The top panels show the resistance switching behavior over time, the middle panels present the dynamically adjusted driving voltage of the MTJ during each calibration step, and the bottom panels illustrate the rolling probability (calculated over the last 50 steps). In both cases, the system starts with randomly initialized voltage conditions, rapidly and automatically converges to the target probabilities, and stabilizes with high precision.}
\end{figure*}

One of the most compelling features of the original self-stabilization strategy is its ability to function without any prior calibration or characterization of the magnetic tunnel junction (MTJ) device. Unlike traditional random number generation systems, which often require an initial phase of tuning or statistical preconditioning to ensure correct operation, the self-stabilized approach enables the generator to begin operating from any arbitrary voltage and converge automatically to the desired switching probability. This property not only simplifies the system architecture but also enhances adaptability and robustness, particularly in scenarios where device parameters vary over time or from chip to chip.

This ability to self-correct is rooted in the algorithm's feedback design: the write voltage is continuously adjusted based on the observed bitstream, nudging the switching probability toward the target value regardless of the starting conditions. As shown in earlier sections, this mechanism is effective at suppressing drift, but when combined with the Downcalibration-2 strategy—where voltage adjustments are applied once every two steps and no bits are discarded—it also preserves full randomness quality. The combination retains the key advantage of automatic convergence while eliminating the correlation artifacts that plagued the original step-by-step implementation.

FIG.~\ref{fig:twoeight} illustrates the automatic convergence capability of the Downcalibration-2 strategy under two extreme cases: targeting switching probabilities of 20\% (panel a) and 80\% (panel b). In both experiments, the system starts from a completely random initial voltage—one that is neither pre-characterized nor pre-optimized. The top panels in each subplot show the resistance switching behavior over time, where the stochastic nature of the MTJ's state transitions is evident. The middle panels display the corresponding voltage values applied at each calibration step. Initially, the voltage values fluctuate widely as the feedback loop begins to adjust from a poorly tuned starting point. However, within a relatively short number of iterations, the voltage converges to a stable range that consistently produces the desired switching behavior. The most informative result appears in the bottom panels, which plot the rolling probability calculated over the last 50 steps. In both the 20\% and 80\% target cases, the system exhibits a clear convergence trajectory: starting from an arbitrary switching probability (which may initially be far from the target), the rolling average rapidly approaches the correct value. Once the convergence is achieved, the switching probability remains stably locked around the target with minimal fluctuation. This demonstrates not only the effectiveness of the control algorithm but also its independence from initial conditions—a crucial property for practical deployment.

The significance of this behavior is substantial. In many real-world applications, such as embedded systems, cryptographic hardware, or field-deployed sensors, pre-characterization of each individual TRNG instance may be impractical or even impossible. Environmental factors like temperature, supply voltage variation, or device aging can further complicate the use of static calibration methods. A self-stabilizing system that is capable of automatically adapting to its own switching characteristics in situ offers a dramatically more robust and scalable solution. The Downcalibration-2 strategy preserves this flexibility while simultaneously delivering high statistical quality and operational efficiency.

Moreover, this automatic convergence also supports configurability. The target switching probability is not hardcoded into the device but can be specified dynamically, allowing the same hardware to serve different purposes—for example, generating biased random streams for stochastic computing or probabilistic inference applications. By simply updating the target probability parameter \(P_0\), the same convergence mechanism will re-tune the voltage until the new target is reached. This reconfigurability, combined with drift suppression and zero-overhead initialization, makes the Downcalibration-2 strategy a highly versatile core for modern TRNG systems.

While the feedback loop exhibits fast convergence within several hundred cycles under most scenarios, it does require a transient period to stabilize after abrupt environmental changes or system startup. In time-sensitive applications such as probabilistic computing, this initial delay may slightly increase the time-to-solution. However, for many long-running applications or streaming architectures, this one-time convergence cost is amortized over the operation duration. In practical hardware, convergence speed can be further improved through techniques such as initialization based on previous history, parallel device arrays, or pipelined operation with dynamic masking of transient bits.

\subsection{Pulse Width Strategy}

\subsubsection{Amplitude vs Width}

\begin{figure*}
\includegraphics[width=\textwidth]{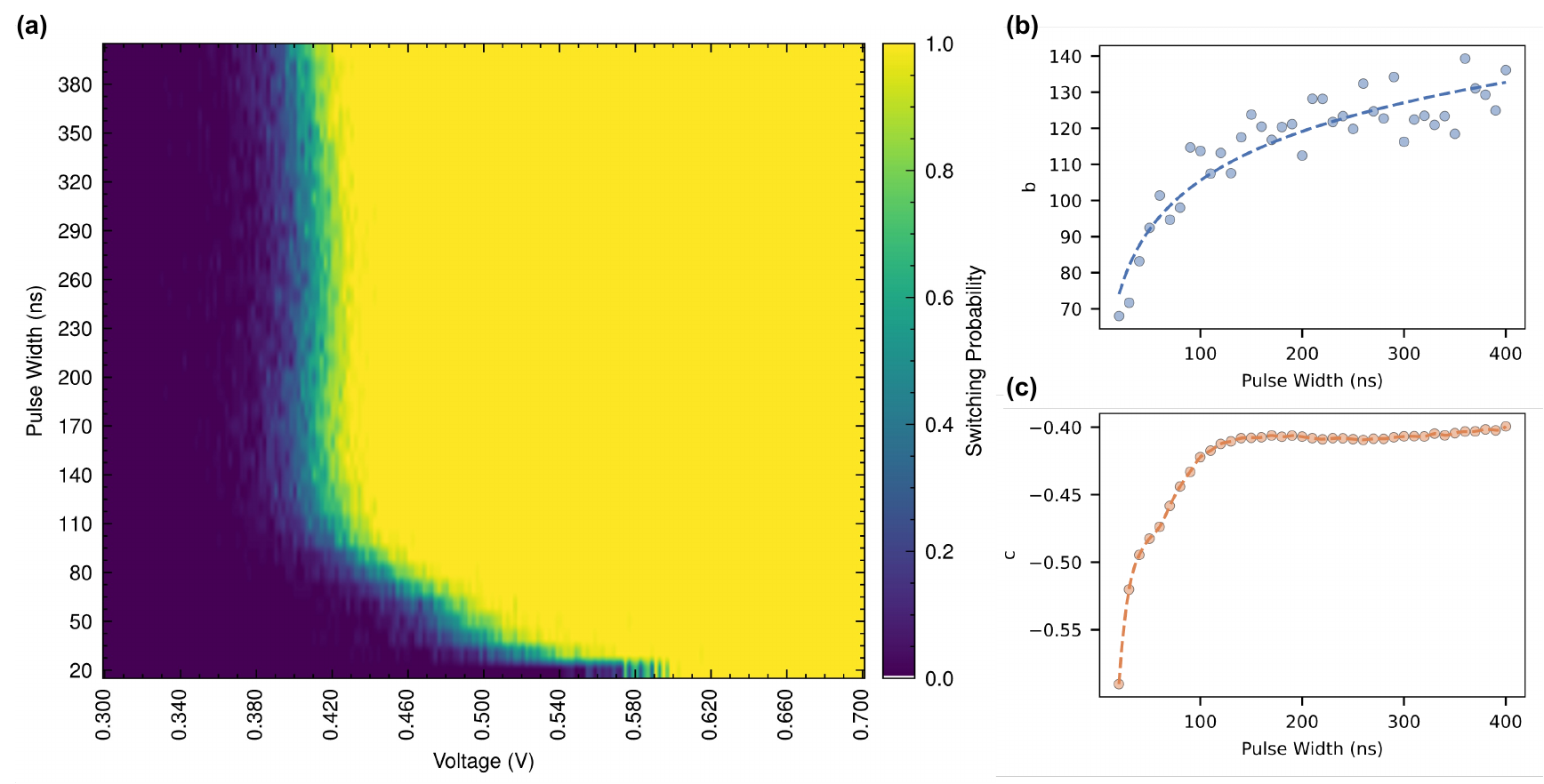}
\caption{\label{fig:pulsetwod}Dependence of switching probability on pulse amplitude and pulse width for the MTJ-based random number generator. (a) Switching probability as a function of voltage amplitude and pulse width. Measurements were conducted over a voltage range of 0.3–0.7 V and pulse widths from 20 to 400 ns. The colormap represents the switching probability, with yellow indicating a probability close to 1 and purple representing a probability close to 0. The transition region reflects the voltage and pulse width combinations where the switching probability is tunable. (b) Relationship between the sigmoid function parameter $b$ and pulse width. The sigmoid function is defined as \( y = 1/{1 + \exp(-b \cdot (x - c))} \). Parameter $b$, representing the slope of the sigmoid curve, increases with pulse width. The dashed line shows a logarithmic fit, indicating the trend. (c) Dependence of the sigmoid function parameter $c$ on pulse width. Parameter $c$, corresponding to the negative value of the voltage at which the switching probability equals 50\%, approaches saturation with increasing pulse width. The dashed curve illustrates the fitting trend.}
\end{figure*}

While the self-stabilization process improves the stability of the switching probability by adjusting the voltage, we also acknowledge that in practical applications, controlling the voltage amplitude is not always a straightforward task, especially when dealing with hardware constraints and high-speed requirements. Voltage adjustments may require significant precision and could be challenging in digital systems where fine control over voltage is not available or is difficult to implement via complex analog-to-digital and digital-to-analog conversions. This limitation points toward an alternative approach for improving the switching probability stability: adjusting the pulse width instead of the voltage. In this section, we explore the dependence of the switching probability on both the voltage amplitude and pulse width, as shown in FIG.~\hyperref[fig:pulsetwod]{\ref{fig:pulsetwod}}. Specifically, we focus on pulse width because it is more practical in real-world applications, where pulse width modulation (PWM) or other time-based control methods can be easily implemented by system clocks. FIG.~\hyperref[fig:pulsetwod]{\ref{fig:pulsetwod}(a)} illustrates the relationship between the switching probability and both the voltage amplitude and pulse width. Measurements were conducted over a voltage range from 0.3 V to 0.7 V and pulse widths ranging from 20 ns to 400 ns. The colormap in this figure represents the switching probability, with yellow indicating a high probability near 1 and purple representing a low probability near 0. The transition region between these extremes is particularly interesting, as it shows the range of voltage and pulse width combinations where the switching probability is tunable. This region allows for precise control over the switching behavior of the MTJ, making pulse width a viable and practical parameter for tuning the switching probability in real-world applications.

In addition to the voltage amplitude, pulse width plays a crucial role in controlling the switching behavior. To better understand this relationship, we analyze the dependence of the switching probability on pulse width through a sigmoid function. As shown in FIG.~\hyperref[fig:pulsetwod]{\ref{fig:pulsetwod}(b)}, we plot the sigmoid function parameter $b$, which represents the slope of the curve, as a function of pulse width. The sigmoid function itself is given by:

\begin{equation}
y = \frac{1}{1 + \exp(-b \cdot (x - c))}
\end{equation}

where $x$ is the pulse width, $y$ is the switching probability, and $b$ and $c$ are parameters that describe the shape of the curve. FIG.~\hyperref[fig:pulsetwod]{\ref{fig:pulsetwod}(b)} shows that the parameter $b$, which determines the steepness of the curve, increases with pulse width. This suggests that as the pulse width decreases, the transition from low to high switching probabilities becomes smoother, allowing for more precise control over the switching probability. The dashed line in the figure shows a logarithmic fit to the data, indicating that the increase in $b$ with pulse width follows a predictable trend. FIG.~\hyperref[fig:pulsetwod]{\ref{fig:pulsetwod}(c)} further explores the relationship between pulse width and the parameter $c$, which corresponds to the voltage of opposite sign at which the switching probability reaches 50\%. This value represents the point where the MTJ switches with equal likelihood between the two states (parallel and antiparallel). As shown in the plot, the parameter $c$ approaches saturation with increasing pulse width, meaning that for longer pulse widths, the voltage at which the switching probability is 50\% stabilizes and does not change significantly. The dashed curve in the figure illustrates the fitting trend, demonstrating that the relationship between pulse width and $c$ is well-defined and predictable. The insights provided by FIG.~\hyperref[fig:pulsetwod]{\ref{fig:pulsetwod}} are valuable for understanding how pulse width can be used to control the switching probability in MTJ-based TRNGs. Given that pulse width is easier to control in practical applications compared to voltage amplitude, these findings suggest that pulse width modulation offers a more practical and efficient method for tuning the switching behavior of MTJs. This approach not only enhances the control over the random number generation process but also simplifies the implementation of MTJ-based TRNGs in hardware systems, making them more adaptable to a wide range of applications.

\subsubsection{Sigmoidal Fitting}

\begin{figure}
\includegraphics[width=0.48\textwidth]{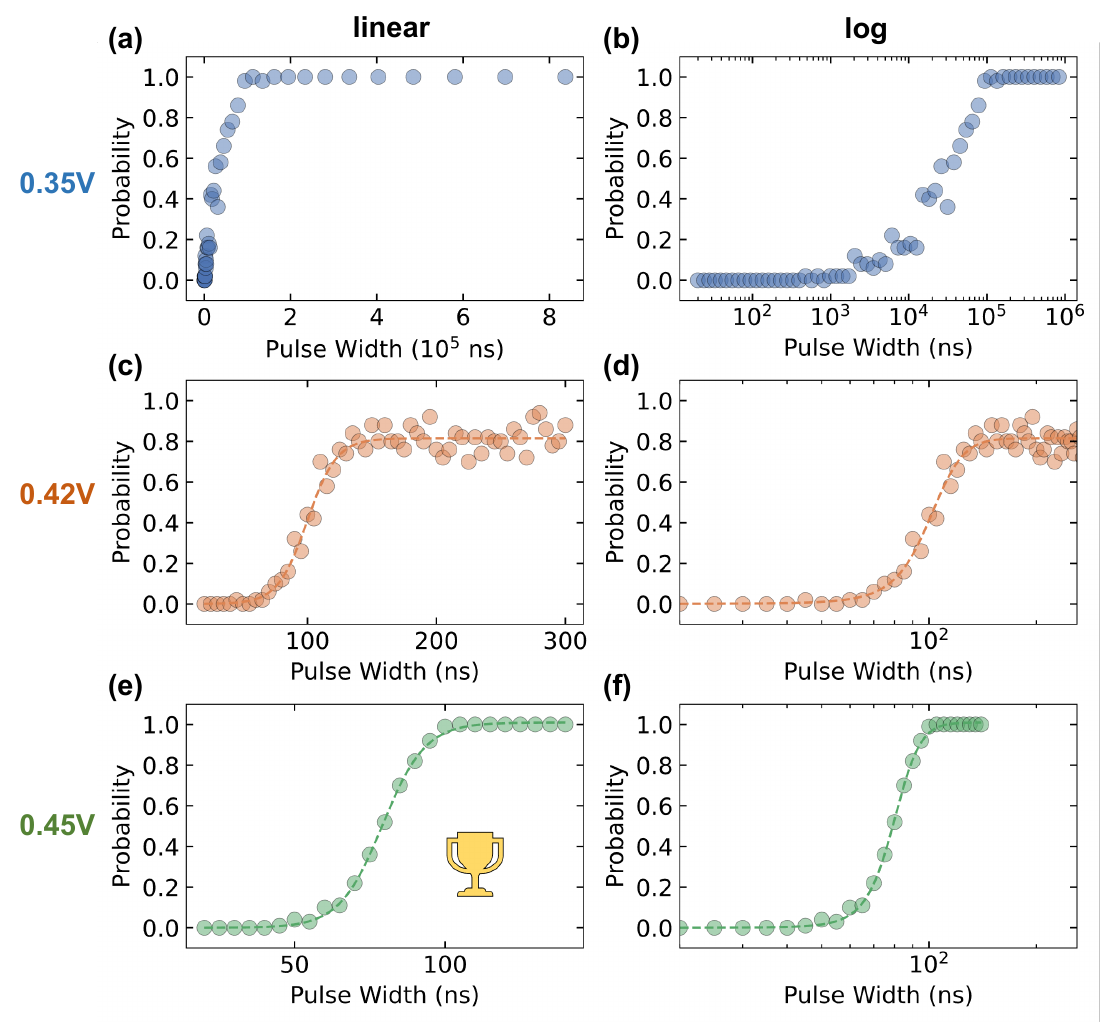}
\caption{\label{fig:fit}Analysis of pulse width dependence on switching probability with Sigmoid fitting. The relationship between pulse width and switching probability is investigated under three fixed voltage conditions (0.35 V, 0.42 V, and 0.45 V). The panels display results in both linear (a, c, e) and logarithmic (b, d, f) scales for pulse width. (a, b) At 0.35 V, the switching probability shows a tunable range across extended pulse widths. However, the transition behavior is irregular and cannot be accurately described using a Sigmoid function, as shown in both linear (a) and logarithmic (b) representations. (c, d) At 0.42 V, the switching probability exhibits a more defined transition region between 50 ns and 150 ns, indicating partial compatibility with Sigmoid fitting. However, the probability saturates prematurely at approximately 80\%, limiting its applicability for certain use cases. (e, f) At 0.45 V, the switching probability demonstrates an ideal Sigmoid-like transition curve, well-described by a sigmoid function across the entire range. The linear scale (e) highlights the smooth transition.}
\end{figure}

\begin{figure*}
\includegraphics[width=0.9\textwidth]{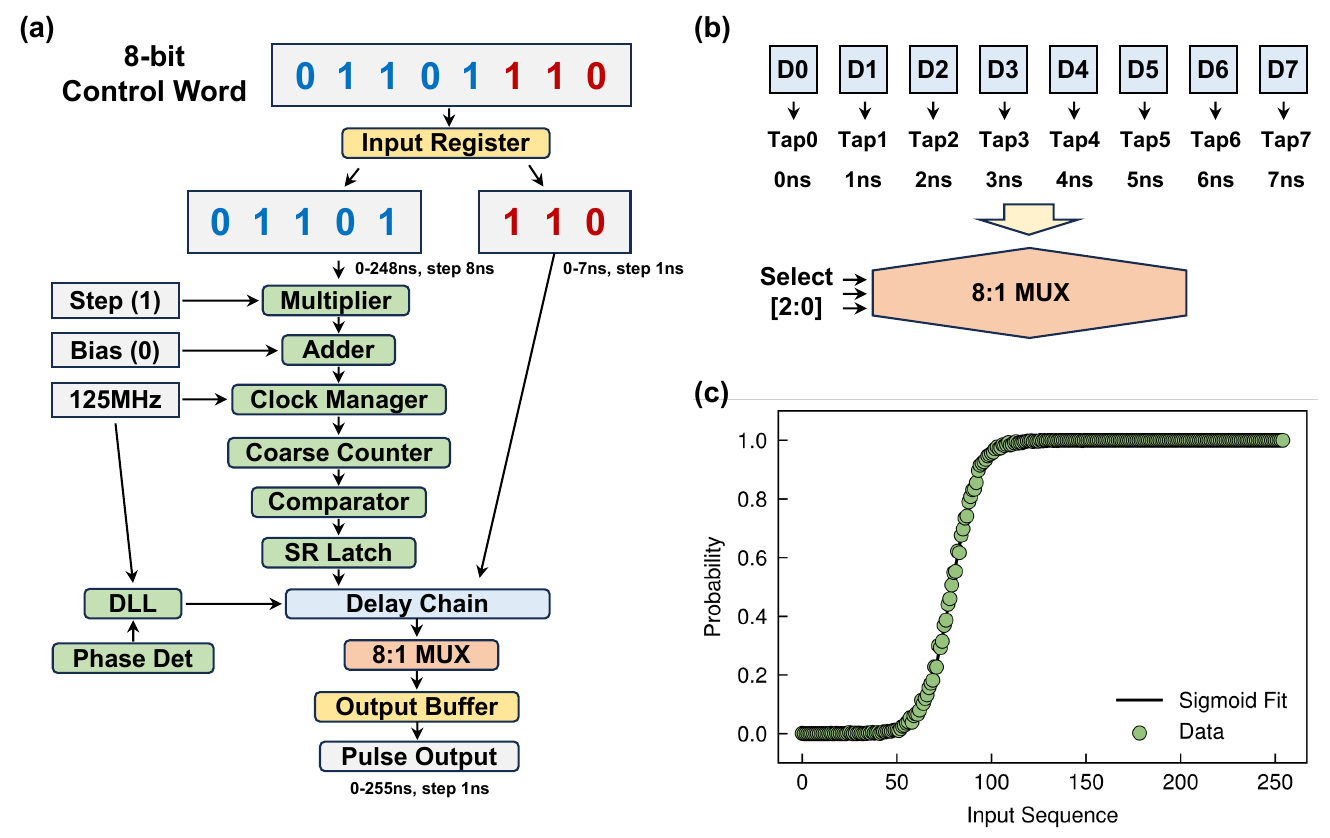}
\caption{\label{fig:circuit}Design and simulation results of a pulse-width-based tunable true random number generator (TRNG). (a) Schematic of the tunable pulse generator driven by an 8-bit control word and a 125 MHz clock, enabling adjustable pulse widths with a 1 ns step size. The upper 5 bits control coarse steps from 8 ns to 255 ns in increments of 8 ns, while the lower 3 bits achieve fine-tuning at 1 ns resolution through a delay chain. (b) Illustration of the 8-bit delay chain and the 8:1 multiplexer (MUX) for selecting precise delay intervals, facilitating the micro-adjustments required for fine-grained control. (c) Simulation results showing the relationship between the 8-bit control word (0–255) and the corresponding switching probabilities. The data points closely match the Sigmoid fitting curve, demonstrating the generator's ability to produce a well-defined probability distribution across the input sequence.}
\end{figure*}

FIG.~\hyperref[fig:fit]{\ref{fig:fit}} investigates the dependence of switching probability on pulse width under three fixed voltage conditions (0.35 V, 0.42 V, and 0.45 V), offering insights into how the pulse width influences the switching behavior of the MTJ-based TRNG at various voltage levels. The results are presented in both linear and logarithmic scales for pulse width, allowing for a comprehensive understanding of the transition characteristics across a wide range of conditions. FIG.~\hyperref[fig:fit]{\ref{fig:fit}(a)} shows the switching probability as a function of pulse width at a fixed voltage of 0.35 V. Over extended pulse widths, the switching probability exhibits a tunable range, indicating that the switching behavior is sensitive to pulse width adjustments. However, the transition between different probability states is irregular, making it difficult to model using a simple Sigmoid function. This irregularity is evident in both linear (FIG.~\hyperref[fig:fit]{\ref{fig:fit}(a)}) and logarithmic (FIG.~\hyperref[fig:fit]{\ref{fig:fit}(b)}) representations, where the transition does not follow the expected smooth curve of a Sigmoid function. This behavior suggests that at this voltage, the MTJ is not operating in a regime where a predictable, continuous relationship between pulse width and switching probability can be established, potentially limiting the precision of random number generation in practical applications. In contrast, at a voltage of 0.42 V (FIG.~\hyperref[fig:fit]{\ref{fig:fit}(c)} and \hyperref[fig:fit]{\ref{fig:fit}(d)}), the switching probability exhibits a more defined transition region. Between pulse widths of 50 ns and 150 ns, the switching probability increases in a more predictable manner, showing partial compatibility with a Sigmoid function. While the transition region is clearer compared to the lower voltage condition, the switching probability saturates prematurely at approximately 80\%, which limits the utility of this voltage range for achieving a broad tunable switching probability. This saturation effect is especially important when a wider range of probabilities is needed, as it prevents the MTJ from reaching higher probabilities, potentially reducing the versatility of the system for generating true random numbers in certain applications.

Finally, at 0.45 V (FIG.~\hyperref[fig:fit]{\ref{fig:fit}(e)} and \hyperref[fig:fit]{\ref{fig:fit}(f)}), the switching probability shows a near-ideal Sigmoid-like transition, with a smooth, continuous curve that fits well across the entire range of pulse widths. This allows for precise control over the switching probability with no premature saturation, making it highly suitable for applications requiring a wide range of random numbers. The linear scale (FIG.~\hyperref[fig:fit]{\ref{fig:fit}(e)}) highlights the smoothness of the transition, demonstrating that at this voltage level, the MTJ-based TRNG operates in a more predictable and reliable regime. The logarithmic scale (FIG.~\hyperref[fig:fit]{\ref{fig:fit}(f)}) further supports this observation, showing a good agreement with a sigmoid fitting curve over the entire range of pulse widths.

\subsubsection{Hardware Implementation and Simulation}

FIG.~\hyperref[fig:circuit]{\ref{fig:circuit}} presents the design and simulation results of a pulse-width-based tunable TRNG and illustrates how precise pulse width control is achieved using an 8-bit control word. The figure highlights the hardware schematic, fine-grained control mechanisms, and the resulting probability distribution, which is well-modeled by a Sigmoid function. Below is a detailed explanation of each component and the associated results. In FIG.~\hyperref[fig:circuit]{\ref{fig:circuit}(a)}, the schematic of the tunable pulse generator is shown, driven by an 8-bit control word and a 125 MHz clock. This generator enables the adjustment of pulse widths with high precision, making it suitable for generating tunable switching probabilities. The 8-bit control word is split into two groups: the upper 5 bits control coarse steps of the pulse width, ranging from 8 ns to 255 ns in increments of 8 ns, while the lower 3 bits provide fine-tuning at a resolution of 1 ns. This division ensures both broad-range configurability and fine-grained precision. The pulse generator is powered by a clock manager that supplies a stable 125 MHz reference clock, and the outputs are further synchronized and optimized by downstream circuitry.

FIG.~\hyperref[fig:circuit]{\ref{fig:circuit}(b)} provides a detailed view of the delay chain and 8:1 multiplexer (MUX) used to implement the fine-tuning mechanism. The 8-bit control word interacts with a delay chain consisting of multiple delay elements, each introducing a precisely controlled delay. The lower 3 bits of the control word are fed into the 8:1 MUX, which selects the desired delay interval from the chain to achieve micro-adjustments in the pulse width. This design allows for precise control over the final output pulse width, ensuring high resolution and reproducibility. The architecture combines digital and analog precision, leveraging the coarse adjustment from the upper 5 bits and the micro-adjustments from the delay chain, to achieve an effective balance between flexibility and accuracy. FIG.~\hyperref[fig:circuit]{\ref{fig:circuit}(c)} shows the simulation results, illustrating the relationship between the 8-bit control word (ranging from 0 to 255) and the corresponding switching probabilities. The data points obtained from the simulation closely align with the Sigmoid fitting curve, demonstrating the effectiveness of the design in producing a well-defined probability distribution across the entire range of input values. This alignment confirms the predictability and consistency of the TRNG's output, with the control word providing direct and precise mapping to switching probabilities. The transition region of the Sigmoid curve is particularly significant, as it represents the range where the switching probability is most sensitive to changes in pulse width, allowing for fine control over random number generation.

\subsection{Hybrid Control Strategies}

\begin{figure*}
\includegraphics[width=0.9\textwidth]{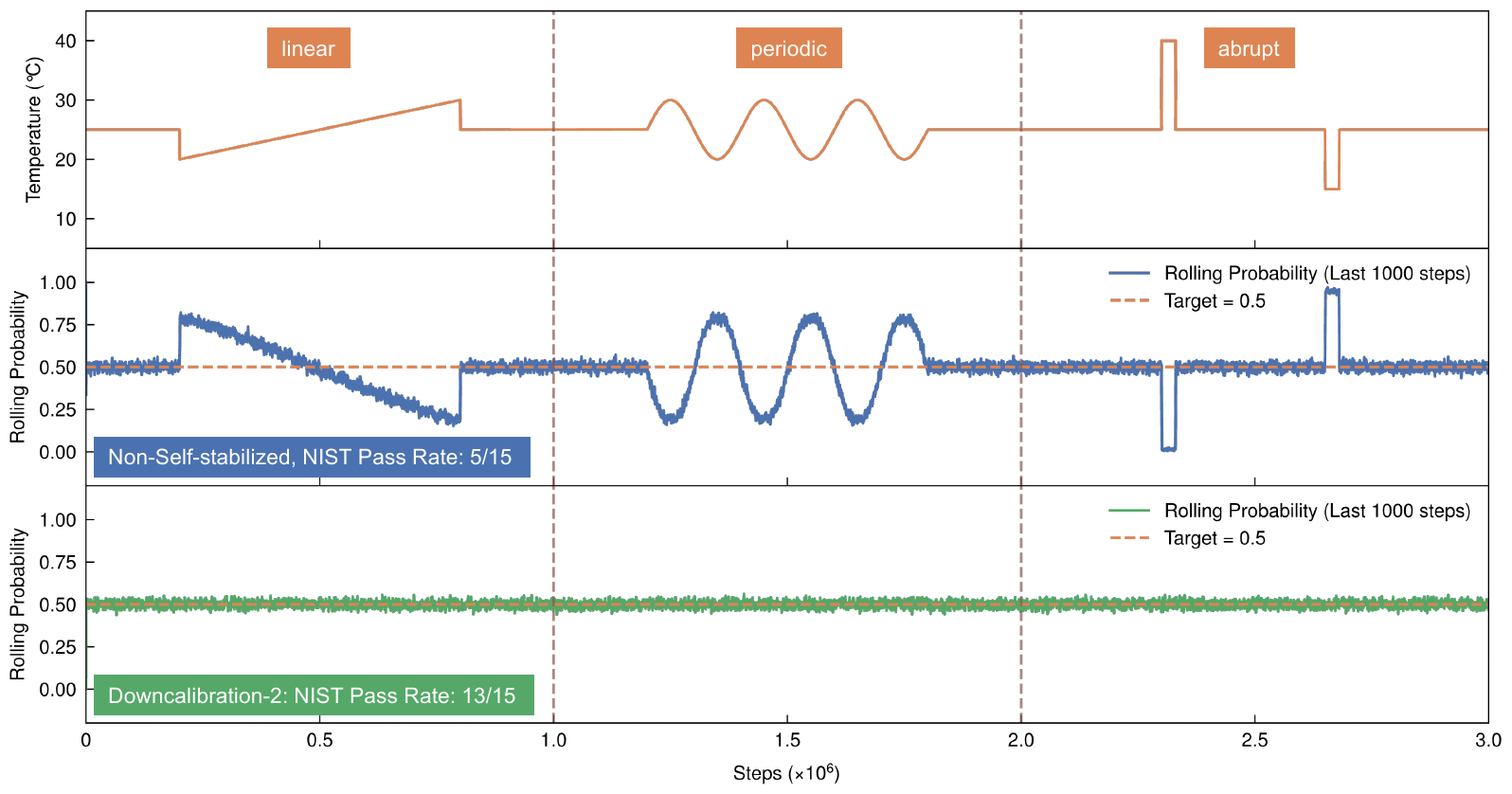}
\caption{\label{fig:sim}Simulated evaluation of random number generator performance under temperature drift using hybrid control strategies. The top panel shows the simulated temperature profile over three consecutive segments of one million steps each, exhibiting linear drift, periodic variation, and abrupt changes. The middle and bottom panels display the rolling switching probability (calculated over the last 1000 steps) under the Non-Self-stabilized and Downcalibration-2 strategies, respectively. The Downcalibration-2 strategy effectively suppresses drift-induced deviation, maintaining stable output close to the target probability of 0.5. It also shows a significant improvement in randomness quality, passing 13 out of 15 NIST SP 800-22 tests, compared to only 5 passed tests for the Non-Self-stabilized case.}
\end{figure*}

The previous sections have established that the Downcalibration-2 self- stabilized strategy, when paired with either voltage or pulse width control, is capable of maintaining statistically robust output while suppressing short-range correlations and long-term drift. However, real-world operating environments introduce additional challenges—particularly thermal fluctuations—that can severely impact device behavior. The switching probability of MTJs is known to be highly sensitive to temperature, as elevated temperatures reduce the energy barrier for switching, shifting the critical voltage or pulse width required to maintain a specific switching probability. Therefore, it is crucial to assess whether the proposed hybrid control framework remains effective under dynamically changing thermal conditions.

To this end, we conducted a comprehensive simulation to evaluate the robustness of the proposed strategies under complex temperature drift profiles. As illustrated in Fig.~\ref{fig:sim}, the system is subjected to three distinct temperature variation regimes, each lasting one million time steps: a linear temperature drift, a periodic oscillation, and a sudden temperature jump. These three segments emulate typical thermal behaviors in embedded systems or field-deployed devices—namely, gradual heating due to sustained operation, ambient temperature cycling, and abrupt environmental transitions. The top panel of Fig.~\ref{fig:sim} shows the synthesized temperature trajectory, where each regime clearly demonstrates the corresponding temperature trend.

The middle panel of Fig.~\ref{fig:sim} presents the performance of the random number generator using a non-self-stabilized approach, in which the pulse width remains constant and no feedback mechanism is used to track or compensate for the changing temperature. The rolling switching probability, calculated over the most recent 1000 output bits, exhibits pronounced fluctuations across all three segments. During the linear drift phase, the probability steadily deviates from the target value of 0.5; in the periodic segment, the output mirrors the thermal oscillations, resulting in sinusoidal fluctuations of the bitstream's bias; and in the abrupt change region, the system fails to recover quickly, producing a sustained deviation from the desired statistical properties. Additionally, the bitstream generated in this configuration passes only 5 out of 15 tests in the NIST SP 800-22 statistical suite, indicating serious degradation in randomness quality due to unmitigated temperature drift.

In contrast, the bottom panel of Fig.~\ref{fig:sim} displays the system's performance under the Downcalibration-2 strategy using pulse width as the tuning parameter. In this implementation, the pulse width is updated every two steps based on the most recent bit outcome, with adjustments quantized to the nearest integer nanosecond. This choice reflects a realistic hardware constraint, as pulse-width control is typically achieved through digital clock logic with limited timing granularity. Remarkably, even with such coarse resolution, the system successfully tracks all three types of thermal variation—linear, periodic, and abrupt. The rolling switching probability remains centered around the target value of 0.5, with only transient deviations observed at the transitions between segments, demonstrating excellent adaptive stability. To further evaluate the quality of the generated random bitstream, the full dataset was subjected to the NIST SP 800-22 statistical test suite. The hybrid strategy resulted in a pass rate of 13 out of 15 tests, a significant improvement compared to only 5 tests passed in the non-stabilized case. The two tests that failed were the Universal Statistical Test and the Approximate Entropy Test. Both of these tests are particularly sensitive to long-range patterns or subtle compressibility in the bitstream. We speculate that these failures are not due to fundamental weaknesses in the control strategy itself, but rather stem from the specific structure of the simulated thermal profile—particularly the periodic segment. Because the simulation used the entire three-million-bit sequence—including portions corresponding to highly dynamic or structured temperature changes—the resulting feedback regulation may have inadvertently introduced weak periodic components or reduced local entropy during adaptation. In real-world applications, however, such behavior can be mitigated by disregarding data collected during phases of external instability (e.g., thermal transients or startup conditions). In practice, TRNGs are often embedded in systems that monitor their own thermal or operational state, and entropy harvesting can be selectively gated to exclude data from unreliable conditions. Moreover, it is important to recognize that not all use cases demand full compliance with all 15 NIST tests. The growing field of "de-randomization" and entropy extraction provides a variety of robust techniques for transforming imperfect randomness into high-quality output suited for cryptographic or computational use. In many stochastic computing and probabilistic inference applications, a bitstream that is statistically sound and free from strong bias is sufficient—even if it fails certain stringent randomness tests.

These results underscore the key advantages of the hybrid control strategy, particularly when implemented using pulse width modulation with Downcalibration-2. First, the system exhibits excellent adaptability to dynamic and unpredictable environmental conditions without requiring any external temperature monitoring or compensation circuits. Second, it achieves this resilience using only simple digital logic and integer-based control steps (1 ns resolution), avoiding the complexity and noise sensitivity of analog voltage regulation. This makes the approach highly suitable for hardware integration, especially in applications where power efficiency, circuit simplicity, and environmental robustness are paramount.

Furthermore, the strong alignment between the pulse-width-based switching characteristics and the self-correcting feedback loop of Downcalibration-2 ensures long-term statistical stability without the need for recalibration or device-specific tuning. The system begins from arbitrary initial conditions and autonomously adapts to both internal parameter drift and external thermal changes. The ability to preserve entropy and pass rigorous randomness benchmarks under such conditions highlights the strength and generality of the proposed approach.

\section{Conclusion}

A drift-resilient TRNG based on MTJs has been developed through the integration of hybrid control strategies that combine self-stabilizing feedback with pulse width modulation. This engineering-oriented design directly addresses the problem of switching probability drift—caused by temperature fluctuations, training effects, and environmental variability—which often limits the reliability of MTJ-based TRNGs in practical deployments. Although traditional bit-by-bit feedback can suppress drift effectively, it introduces undesirable short-range correlations in the output bitstream. To resolve this issue, a Downcalibration-2 strategy was proposed and validated through extensive simulation. This method updates the control parameter every two steps without discarding any data, thereby achieving both statistical independence and rapid convergence. The approach preserves high entropy and passes all randomness tests defined by the NIST SP 800-22 suite under stable conditions. To enhance compatibility with digital hardware, the control mechanism was shifted from voltage tuning to pulse width modulation. Experimental characterization revealed a well-defined sigmoidal relationship between pulse width and switching probability. A practical 1 ns-resolution pulse generator was designed using an 8-bit control word, enabling precise and reconfigurable probability tuning suitable for standard CMOS integration.

The robustness of the hybrid strategy was further evaluated under simulated temperature drift scenarios, including linear ramping, periodic oscillation, and abrupt fluctuations. The Downcalibration-2 scheme demonstrated strong adaptability to these variations, maintaining output stability without requiring prior calibration or environmental sensing. Such resilience makes the system well-suited for real-world applications, especially those in security, probabilistic computing, and low-power embedded systems.

While the self-stabilizing feedback introduces a transient convergence period, our results show that stability is typically reached within a few hundred cycles. For most use cases, this initial overhead is amortized over longer operation periods. In future designs, warm-start initialization and system-level masking can be employed to further reduce response time. The proposed framework achieves a compelling balance of statistical quality, control simplicity, and implementation feasibility. Although this work focuses on validating the statistical performance and thermal resilience of the hybrid strategy, future investigations will extend to quantitative comparisons against other TRNG technologies, considering bitrate, power efficiency, area footprint, and system-level integration. Such efforts will further establish MTJ-based TRNGs as a practical and scalable solution for next-generation secure and stochastic hardware platforms.

\section*{Data Availability}
All data needed to evaluate the conclusions in the paper are present in the paper or openly available \cite{zhang_data_2025}. Additional data available from authors upon request.

\section*{Author Contributions}
C.W., T.K., and X.H. led and was involved in all aspects of the project. R.Z. was responsible for the conceptualization, execution, and coordination of the study. C.W. and R.Z. conducted the primary experiments and contributed to the design and analysis of the study. S.H., J.G., Q.D., Y.S., and Z.Z. prepared the STT-MTJ samples. R.Z., C.W. and Y.X. performed key measurements and analyses. R.H. and M.H. offered technical support and assisted with specific measurements. R.H., M.H., Y.X., X.L., and T.K. provided critical discussions and contributed to the interpretation of results. All authors were involved in discussions and contributed to the writing and revision of the manuscript.

\begin{acknowledgments}
This work was supported by the National Key Research and Development Program of China (MOST) (Grant No. 2022YFA1402800), the National Natural Science Foundation of China (NSFC) (Grant Nos. 12134017, 51831012, 51620105004, and 12374131), the Strategic Priority Research Program (B) of Chinese Academy of Sciences (CAS) (Grant Nos. XDB33000000). C. H. Wan appreciates financial support from the Youth Innovation Promotion Association, CAS (Grant No. 2020008).
\end{acknowledgments}

\bibliography{references}

\end{document}